\def \half {\frac12}
\def \pa {\partial}
\def \M {{\cal M}}

\def \R {{\it R}}
\def \d {\hbox{d}\,}

\def \e {\hbox{e}}

\def \v {\vskip}

\def \ephi {\e^{-2\phi}}
\def \erho {\e^{2\rho}}
\def \pp  {\partial_+}
\def \pm  {\partial_-}
\def \a {\alpha}
\def \b {\beta}
\def \m {\mu}
\def \n {\nu}
\def \r {\rho}
\def \f {\phi}
\def \d {\delta}
\def \l {\lambda}
\def \o {\omega}

\def \be {\begin{equation}}
\def \ee {\end{equation}}
\def \bfr{\begin{flushright}}
\def \efr{\end{flushright}}
\def \caja{\makebox[5.2cm][1]}

\documentstyle[a4,12pt]{article}
\begin{document}


\pagestyle{empty}
\bfr
\caja{Imperial TP/92-93/37}
\efr
\v 3cm
\begin{center}
{\bf SYMPLECTIC STRUCTURE OF 2D DILATON GRAVITY}
\footnote[2]{Work partially supported by the Comisi\'on Interministerial
de Ciencia y Tecnolog\'\i a.}
\v 0.3 cm
A. Mikovi\'c$^1$ and M. Navarro$^2$
\v 0.3 cm
\end{center}
\noindent 1.- The Blackett Laboratory, Imperial College,
Prince Consort Road, London SW7 2BZ. U.K.
\v 0.3cm
\noindent 2.- Departamento de F\'\i sica Te\'orica y del Cosmos,
Facultad de Ciencias, Universidad de Granada, Campus de Fuentenueva,
18002. Granada. Spain.

\begin{center}{\bf Abstract}
\end{center}
We analyze the symplectic structure of two-dimensional dilaton gravity by
evaluating the symplectic form on the space of classical solutions.
The case when the spatial manifold is compact is studied in detail.
When the matter is absent we find that the reduced phase space is a
two-dimensional cotangent bundle and
determine the Hilbert space of the quantum theory.
In the non-compact case the symplectic form is
not well defined due to an
unresolved ambiguity in the choice of the boundary terms.
\vfil\eject

\setcounter{page}{1}
\pagestyle{plain}

\section{Introduction}

The study of 2d dilaton gravity models has attracted a lot
of interest because they can
serve as toy models for understanding the physics of black holes (for
a review and references see \cite{rev}).
A model proposed by Callan, Giddings, Harvey and Strominger (CGHS)
\cite{cghs} has been an object of lot of
studies, because of its nice properties, of which its classical
solvability is crucial for our approach.
As argued by many authors (see \cite{rev}), semiclassical
quantization schemes of CGHS model are not sufficient in order to understand
the quantum fate of the corresponding 2d black hole. Non-perturbative
quantization schemes were proposed by many authors (see \cite{rev}).
In the canonical quantization approaches (see \cite{mik}), the
knowledge of the phase space of
the theory is crucial.

In the standard canonical approach, determination of the true
(reduced) phase space
becomes a non-trivial task because of the presence of the constraints.
In the covariant phase space approach \cite{crnko}, the parameter space of
non-equivalent solutions of the equations of motion is defined to be
the true phase
space of the theory. Since the classical solutions of the CGHS model
are known, one can study the symplectic structure of the space of the
solutions in order to find the reduced phase space of the theory and
also to find suitable variables for quantization.

This is usually achieved by evaluating the symplectic form on the
space of the solutions, whose definition was given in \cite{crnko}.
This method was employed to study certain 2d gravity models, like
Jackiw-Taitelboim and induced gravity models \cite{nna}. In this paper
we apply the techniques developed in \cite{nna} to the case of the
CGHS model.

\section{General solution.}

The classical action of the CGHS model can be written as
\be S_{CGHS} = \frac{1}{2}\int_{\cal M}\,d^2x\sqrt{-g}
\left[\e^{-2\phi}(R+4(\nabla\phi)^2 +4\lambda^2) -
\frac{1}{2}(\nabla f_i)^2\right]\>,\ee
where $R$ is the scalar curvature corresponding to the 2d metric
$g_{\m\n}$, $\nabla$ is the corresponding covariant derivative,
while $\phi$ and $f_i$, $i=1,...,N$, are scalar fields. $\cal M$ is
the 2d manifold with the topology of $\Sigma\times{\rm R}$. When
$\Sigma = {\rm R}$, the solution of the equations of motion can be
interpreted as a 2d black hole. When $\Sigma = S^1$ (circle), then it is
unclear whether a black hole interpretation is valid, although some
authors tried to argue a relevance in the limit of a ``large'' circle
\cite{{mik},{hirano}}.

In 2d one can always chose the conformal gauge for the metric
\be d s^2 = -\e^{2\rho} d x^+ d x^-\, \ee
or
\be g_{\mu\nu} = -\frac{1}{2}\left(\begin{array}{cc}0&\erho\\
\erho&0\end{array}\right)\ee
where $x^+ = t+x, \>x^-=t-x$.
The equations of motion are then given by \cite{cghs}
\begin{eqnarray}
 T_{++} = \ephi\left(\pp(2\rho)\pp(2\phi) - \pp^2(2\phi)\right)+
\frac{1}{2}\pp f\pp f &=& 0\\
T_{--} = \ephi\left(\pm(2\rho)\pm(2\phi) - \pm^2(2\phi)\right)+
\frac{1}{2}\pm f\pm f &=& 0\\
T_{+-} = \ephi\left(\pp\pm(2\phi)-\pp(2\phi)\pm(2\phi)-
\lambda^2\erho\right)&=&0\label{1.a}\\
-2\pp\pm(2\phi) + \pp(2\phi)\pm(2\phi) +\pp\pm(2\rho)
+\lambda^2\erho &=&0\\
\pp\pm f &=&0\end{eqnarray}
Eqs. (6) and (7) are equivalent to
\begin{eqnarray}
\pp\pm(\ephi) + \lambda^2\e^{(2\rho-2\phi)} &=& 0\label{2.a}\\
\pp\pm(2\rho -2\phi)&=&0\>.\label{3.a}\end{eqnarray}

Eq. (10) implies that
\be \e^{(2\rho -2\phi)} = \pp p \pm m\> \ee
with $p=p(x^+),\>m=m(x^-)$. Eq. (9) then becomes
\be \pp\pm(\ephi) + \lambda^2\pp p \pm m\> = 0\label{5.a}\ee
and can be integrated as
\be \e^{-2\phi} =  -\lambda^2p m +a +b\>, \label{6.a}\ee
where $a=a(x^+)$ and $ b=b(x^-)$.
Thus we can write
\begin{eqnarray}
\e^{2\rho} = \frac{\partial_+p\partial_-m}{-\lambda^2 pm + a + b}
\>.\label{3}\end{eqnarray}

The constrain equations (4) and (5) can be now expressed
in terms of the functions $p,\>m,\>a,\>b$ as

\begin{eqnarray}
0 &=& T_{++} = \left[\partial^2_+a - \frac{\partial^2_+p}{\partial_+p}\pp
a\right]
+\frac{1}{2}\partial_+f\partial_+f\>,\label{5}\\
0 &=& T_{--} = \left[\partial^2_-b - \frac{\partial^2_-m}{\partial_-m}\pm
b\right]
+\frac{1}{2}\partial_-f\partial_-f\>.\label{6}\end{eqnarray}

The general solution of the eqs. (15) and (16) is given by
\begin{eqnarray}
a &=& a_0 +\a p -\half\int^{x^+} d y^+ \pp p \>\int^{y^+}d
z^+\left\{{1\over\pp p} \pp f\pp f\right\}\label{10.a}\\
b &=&b_0 + \b m -\half\int^{x^-}d y^- \pm m \>\int^{y^-}d
z^-\left\{{1\over\pm m} \pm f\pm f\right\}\label{11.a}
\end{eqnarray}
where $a_0, b_0, \a, \b$ are arbitrary constants. One can get rid off
the $\a$ and $\b$ terms by redefining $p$ and $m$ by constant shifts.
The CGHS parametrization is obtained by defining $\pp p = e^{w_+}$
and $\pm m = e^{w_-}$ \cite{cghs}.

Therefore the general solution of the system of eqs. (4-8) can be written
as
\begin{eqnarray}
\e^{-2\phi} &=& {M\over\l}-\lambda^2 pm
-\int^{x^+}d y^+\pp p\>\int^{y^+}d
z^+\left\{\frac{1}{\pp p}\frac{1}{2}\pp f\pp f\right\}\nonumber\\
&-&\int^{x^-}d y^-\pm m\>\int^{y^-}d
z^-\left\{\frac{1}{\pm m}\frac{1}{2}\pm f\pm f\right\}\label{14.a}\\
\e^{2\rho} &=& (\partial_+p\partial_-m)\e^{2\phi}\label{15.a}\\
f_i &=& f_i^+(x^+) + f_i^-(x^-)\quad,\label{16.a}\end{eqnarray}
where ${M\over\l}=a_0 + b_0$.
In the case when there is no matter one gets
\begin{eqnarray}
\ephi &=& -\lambda^2pm  + \frac{M}{\lambda} \label{17.a}\\
\erho &=& \frac{\pp p \pm m}
{-\lambda^2pm  + \frac{M}{\lambda} }\, .\label{18.a}\end{eqnarray}

The space of the solutions (19-21) is invariant under the conformal
transformations
\be x^+ = x^+(y^+) \quad,\quad x^- = x^-(y^-)\ee
\be \f (x^+,x^-)= \tilde{\f}(y^+,y^-)\quad,\quad
\rho (x^+,x^-) = \tilde{\r} (y^+,y^-)+ \half {\rm ln} {\pa x^+\over \pa y^+}
{\pa x^-\over \pa y^-}\, .\ee
This residual symmetry can be fixed by specifying the functions $p$
and $m$. This choice depends on the topology of $\M$, and for the time
being we will leave $p$ and $m$ unspecified. In the non-compact case,
the choice $p=x^+$ and $m=x^-$ corresponds to a 2d black hole,
whose mass is given by the parameter $M$ \cite{cghs}.

\section{Symplectic form}

Given a field theory with fields $\Psi^\alpha$ and an action
$S[\Psi]$, one can obtain
the corresponding symplectic form as
\be \omega =
\int_\Sigma d\sigma_\m(-\delta j^\m)\label{defsf}\ee
where $\Sigma$ is a spatial
hypersurface, while $j^\m$ is the symplectic current. $j^\m$ can be
obtained from the variation of the action
\be \delta S =\int \partial_\mu j^\mu +  \int {\d S\over \d \Psi^\a}\delta
\Psi^\alpha  \label{variationofS}\quad.\ee
The symbol $\d$ in (\ref{defsf}) is an
infinite-dimensional generalization of the usual exterior derivative
\cite{crnko}, and it coincides with the usual field variation.
The form $\o$ is conserved in time if the equations of motion are satisfied.
We chose $\Sigma$
to be a $t=const.$ hypersurface. With this choice of $\Sigma$ we have
\be \omega = \int_{t=t_0} d x (-\delta j^0)\label{34.a}\ee
where $j^0 = j^+ + j^-$.

In order to calculate the symplectic form of the
theory (1) we can start from the lagrangian
in the conformal gauge. We get
\be
{\cal L}= \ephi\left(\pp\pm(2\rho) -
\pp(2\phi)\pm(2\phi) +\lambda^2\erho\right)
+\frac{1}{2}\sum_i^N\pp f_i\pm f_i\>.\label{19.a}\ee

Variation with respect to $\phi$ and $\rho$ gives
\begin{eqnarray}
\delta{\cal L} &=& {\d S\over\d\phi} \d\f
+ {\d S\over\d\rho}\d\r\nonumber\\
&+&\pp\left[\ephi\pm\delta(2\rho) -\ephi\pm(2\phi)\delta(2\phi)
+\frac{1}{2}\sum_i\pm f_i \delta f_i\right]\nonumber\\
&+&\pm\left[-\pp(\ephi)\delta(2\rho)-\ephi\pp(2\phi)\delta(2\phi)
+\frac{1}{2}\sum_i\pp f_i\delta f_i\right]\>.\end{eqnarray}

Thus the symplectic current is
\begin{eqnarray}\jmath^+&=&
\ephi\pm\delta(2\rho) -\ephi\pm(2\phi)\delta(2\phi)
+\frac{1}{2}\sum_i\pm f_i \delta f_i\label{30.a}\\
\jmath^-&=& -\pp(\ephi)\delta(2\rho)-\ephi\pp(2\phi)\delta(2\phi)
+\frac{1}{2}\sum_i\pp f_i\delta f_i\>.\label{31.a}\end{eqnarray}

After a long but straightforward computation it can be shown that
the symplectic form takes the following form in terms of the phase
space coordinates:
\be
\omega =\int_{t=t_0} d x \,\omega^0
\ee
where
\begin{eqnarray} \omega^0 &=& \left[\partial_+ -\partial_-\right]
\left\{\delta(-\lambda^2pm +a)\delta\ln \partial_-m +
\delta(-\lambda^2pm)\delta\ln p + \delta a \delta \ln
\partial_+p\right.\nonumber\\
&-& \left.\delta a\delta\ln \partial_+a + \delta b \delta \ln \partial_-b
\right\}\nonumber\\
&+&\frac{1}{2}\left\{-\delta a\delta
\left[\frac{1}{\partial_+a}\partial_+f_i\partial_+f_i\right]
-\delta b\delta
\left[\frac{1}{\partial_-b}\partial_-f_i\partial_-f_i\right]\nonumber\right.\\
&+&\left.\half\left[\delta f_i\delta \partial_-f_i
+\delta f_i\delta \partial_+f_i\right]\right\}\>.\label{8}\end{eqnarray}
It can be seen from (\ref{8})
that the only contribution of the chiral functions $p, m$ is
through boundary terms.

By making use of the constraints (\ref{5}, \ref{6}) and the equality
\begin{eqnarray}
&-&\delta a \delta
\left(\frac{\partial^2_+p}{\partial_+p}-\frac{\partial^2_+a
}{\partial_+a} \right)
-\delta b \delta
\left(\frac{\partial^2_-m}{\partial_-m}-\frac{\partial^2_-b}{\partial_-b}\right)
\nonumber\\=
&-&(\partial_+ - \partial_-)\left[\delta a\delta (\ln \partial_+p -\ln
\partial_+a)\right]
+ \delta \partial_+a\delta (\ln \partial_+p - \ln \partial_+a)\nonumber\\
&-&(\partial_+ - \partial_-)\left[\delta b\delta (\ln \partial_-m -\ln
\partial_-b)\right]
+ \delta \partial_-b\delta (\ln\partial_-m - \ln
\partial_-b)\label{11}\end{eqnarray}
we find a simpler expression for the symplectic form
\begin{eqnarray}
\omega &=& \left[\delta(-\lambda^2pm +a +b)\delta\ln \partial_-m
+\delta(-\lambda^2pm)\delta\ln p\right]_{x_1}^{x_2}\nonumber\\
&+&\int_{x_1}^{x_2}d x \{\delta \partial_-b\delta \ln \partial_-m +
\delta \partial_+a\delta\ln \partial_+p \nonumber\\
 &+&\frac{1}{2}(\delta f_i\delta\partial_-f_i
+\delta f_i\delta\partial_+f_i) \}\>.\label{12}
\end{eqnarray}
which we rewrite as
\begin{eqnarray}
\omega &=& \left[\delta(a +b)\delta\ln \partial_{-}m
+\delta(-\lambda^2 pm)\delta\ln (p\pm m)\right]_{x_1}^{x_2}\nonumber\\
&+&\int_{x_1}^{x_2}d x \{\delta \partial_-b\delta \ln \partial_-m +
\delta \partial_+a\delta\ln \partial_+p \}\nonumber\\
 &+&\frac{1}{2}\int_{x_1}^{x_2} dx \delta f_i\delta\dot{f}_i
,\label{13}
\end{eqnarray}
where $x_{1,2}$ are the endpoints of the $x$-interval, $\dot f = {\pa
f\over \pa t}$ and
\be [F(x)]_{x_1}^{x_2} = F(x_2)-F(x_1)\quad. \ee

Note that the expression (\ref{13}) is not symmetric under the
exchange of $x^+$ and $x^-$, although the space of the solutions is. A
symmetric expression can be achieved by adding a total derivative of a two
form $\pa_x \tilde{\o}$ to $\o^0$. This does not change the symplectic
form in the compact
case. By taking $\tilde{\o}=-\half\d e^{-2\f}\d (2\r - 2\f)$ we get
\begin{eqnarray}
\omega &=& -\half\left[\delta(a +b)\delta\ln{\pp p\over \partial_{-}m}
+\delta(-\lambda^2 pm)\delta\ln{\pp p m\over p\pm m
}\right]_{x_1}^{x_2} \nonumber\\
&+&\int_{x_1}^{x_2}d x \{\delta \partial_-b\delta \ln \partial_-m +
\delta \partial_+a\delta\ln \partial_+p \}\nonumber\\
 &+&\frac{1}{2}\int_{x_1}^{x_2} dx \delta f_i\delta\dot{f}_i
\,.\label{14}
\end{eqnarray}
However, we will use the eq. (\ref{13}) since it is simpler for calculations.

\section{ CGHS model on a circle}

Untill now the discussion of the CGHS model has been general, without
specifying the topology of the manifold where it is defined. In
this section we will chose the 2-dimensional spacetime
manifold to be of the form $S^1\times{\rm R}$.

\subsection{Pure dilaton gravity}

Let us consider first the case without matter, i.e. $f_i = 0$, for all $i$.
The solution for the metric and the dilaton field is then given
by
\begin{eqnarray}
\ephi &=& -\lambda^2pm + \frac{M}{\lambda} \nonumber\\
\erho &=& \frac{\pp p \pm m}
{-\lambda^2pm + \frac{M}{\lambda}
}\label{1.c}\end{eqnarray}

In this case the symplectic form becomes
\be \omega = \int_x^{x+2\pi} (\pp - \pm) W\>.\label{W-W}\ee
Since $\pp -\pm = \partial_x$ we have
\be \omega = W(x+2\pi) -W(x)\>,\label{3.c}\ee
where
\be W =\delta(
 \frac{M}{\lambda})\delta\ln\pm m +
\delta(-\lambda^2 pm )\delta \ln (p\pm m ).\label{W}\ee

It is important to notice here that $p$ and $ m$ do not have to be
singlevalued functions on the circle. They can have a nontrivial
monodromy transformation
\begin{eqnarray} p(x^+ +2\pi) &=& qp(x^+)\nonumber\\
m(x^- -2\pi) &=& \frac{1}{q}m(x^-)\, ,\label{mndr}\end{eqnarray}
where $q$ is an arbitrary positive real number.
Hence $W$ is not a singlevalued two-form on the
circle and therefore $\omega$ is not zero. $\o$ should
be independent of the arbitrarily chossen point $x$, and hence it
should be independent of the functions $p$ and $m$. Indeed this
is the case, which can be seen by inserting (40) into (38)
so that
\be \omega = -\delta\frac{M}{\lambda}\delta\ln q\>.\label{7.c}\ee

We conclude that in the compact case without matter the CGHS model
has no local degrees of freedom. In fact it
behaves like a mechanical sistem with one degree of freedom.
The parameter $M$ behaves like a momentum conjugate to the monodromy parameter
$q$. It is tempting to identify $M$ with the mass of the black hole.
However, it is not clear whether in the compact case such an
identification makes any sense.

Our task now is to determine the exact reduced phase space of the model.
To do that we have to find out the maximal set of functions $p$ and $m$
which, beside fulfiling the right monodromy
transformation properties,
define physical fields with the appropiate sign. In the present case we need
to find out the functions $p$, $m$, with the appropiate monodromy properties
such that both $\e^{2\rho}$ and $\e^{-2\phi}$ are positive everywhere and.
If we take $q =\e^r$ with $r\in {\rm R}$, then any
functions with the appropiate monodromy transformation properties can be
written as:
\begin{eqnarray}
   p &=& \frac{M}{\lambda}\e^{\frac{r}{2\pi}x^+}u\>, \label{pfree}\\
   m &=&\frac{1}{\lambda^2}\e^{\frac{r}{2\pi}x^-}v\>, \label{mfree}
\end{eqnarray}
where $u=u(x^+)$ and $v=v(x^-)$ are periodic functions on the real line.
With this choice for $p$ and $m$, the physical fields take the
form
\begin{eqnarray}
\e^{-2\phi} &=& \frac{M}{\lambda}(1-uv\e^{\frac{r}{2\pi}2t})\>,
\label{phifree}\\
\e^{2\rho} &=&
\frac{\frac{M}{\lambda}\frac{1}{\lambda^2}
\e^{\frac{r}{2\pi}2t}(\pp u +\frac{r}{2\pi}u)(\pm v +\frac{r}{2\pi}v)}
{\frac{M}{\lambda}(1-uv\e^{\frac{r}{2\pi}2t})}\>.\label{rhofree}
\end{eqnarray}
The periodicity of the functions $u$ and $v$,
together with the positivity of the exponential
$\e^{\frac{r}{2\pi}2t}$ and the fact that the coordinates
$x$ and $t$, or what is almost the same $x^+$ and $x^-$, can be varied
independently of each other, makes it easy to realize that the positivity of
both the metric and the exponential of the dilaton field
imply that the parameter $M$ and the
product $uv$ are restricted to be positive (we are taking $\lambda>0$).
The positivity of the product $uv$ restricts in turn the allowed values for
$r$ and $t$. It is clear from the expression (\ref{phifree})
that the range of their allowed values does not cover the entire real line.
For example, if we permit $t$ to be positive then $r$ is bounded by above,
and $r$ is bounded by below if $t$ is negative.

        To go on with the analysis let us consider
the solutions above after
an appropiate gauge choice has been done. A  choice of gauge that fulfils
all the requirements above for $u$ and $v$ is $u=1=v$.
In this gauge, the physical fields take the form:
\begin{eqnarray}
\e^{-2\phi} &=& \frac{M}{\lambda}(1-\e^{\frac{r}{2\pi}2t})\>,
\label{fi2}\\
\e^{2\rho} &=&
\frac{1}{\lambda^2}\frac{(\frac{r}{2\pi})^2\e^{\frac{r}{2\pi}2t}}
{(1 - \e^{\frac{r}{2\pi}2t})}\>.\label{ro2}
\end{eqnarray}

It is clear from (\ref{fi2},\ref{ro2}) that,
for $M>0$, $e^{-2\phi}$ and $e^{2\rho}$ are positive if and only if
$r<0$ and $t>0$ or $r>0$ and $t<0$. The reduced phase space splits
then into two disjoint pieces, one with $r>0$ and the other with $r<0$.

By making a canonical change of variables,
it is possible to rewrite the symplectic form in a way
that makes clear the
cotangent bundle nature of the reduced phase space.
For let us introduce new coordinates $p, s$ defined by
$\frac{M}{\lambda}=\e^p$, $h=r\e^p$. In this coordinates the symplectic
form reads

\be \omega = \delta h\,\delta p\>. \ee

Since the range of $p$ covers the entire real line, it is clear
that the reduced phase space for the free theory and
compact spacelike sections is given by

\be T^*{\rm R_+}\cup T^*{\rm R_-}
\label{freephasespace}\ee

The cotangent bundle nature of the reduced phase space makes it possible to
determine the Hilbert space of the quantum theory. General principles of
the quantum theory indicate that the Hilbert space ${\cal H}$
is given by the square
integrable functions on the configuration space. We have then

\be {\cal H}=L^2({\rm R_+}, \frac{\hbox{d} h}{h})\oplus
L^2({\rm R_-}, \frac{\hbox{d} h}{h})\>,\ee
where the measure $\frac{\hbox{d} h}{h}$ accounts for the restriction in
sign of the parameter $h$.

\subsection{Inclussion of the matter}

In the case when matter fields are present
the symplectic form is given by the eq. (\ref{omega}) with $p$ and $m$
obeying the monodromy transformations (\ref{mndr}), while $a+b$ and
$f^+ + f^-$ are periodic functions.
By procceeding in the same fashion as in the pure dilaton gravity case
we arrive to
\begin{eqnarray}
\omega =&-&\delta(a(x) +b(x))\delta\ln q\nonumber\\
&+&\int_{x}^{x+2\pi} d y\{\delta \partial_-b\delta \ln \partial_-m +
\delta \partial_+a\delta\ln \partial_+p\}\nonumber\\
 &+&\frac{1}{2}\int_{x}^{x+2\pi} dy \delta f_i\delta\dot{f}_i\>. \label{20.c}
\end{eqnarray}
Once again $\omega$ should be independent of $x$. This can
be checked by taking the total derivative with
respect to $x$. The explicit dependence on $x$ of
the boundary term compensates with the nonperiodicity of the
integrand in (\ref{20.c}).

Let us now make a change of variables in order to separate
the monodromy part of the functions  from the parts which
are periodic:

\begin{eqnarray} \pp p &= &\e^{\frac{r}{2\pi}x^+}w_+\>,\label{changep}\\
\pm m &=& \e^{\frac{r}{2\pi}x^-}w_-\>, \label{changem}\end{eqnarray}
with  $q = \e^r$ in order that $p$ and $m$ fulfil the monodromy
transformation properties (\ref{mndr}). Note that
(\ref{changem}-\ref{changep}) is a change of variables that
does not imply any gauge fixing.

We then have
\begin{eqnarray}
\omega = &-&\delta(a(x) +b(x))\delta\ln q\nonumber\\
&+&\int_x^{x+2\pi} d y\left\{\delta \pp b
\left[\delta \ln w_- + \delta\frac{r}{2\pi}y^-\right]\right.\nonumber\\
&+& \delta \partial_+a\delta
\left.\left[\delta \ln w_+ + \delta\frac{r}{2\pi}y^+\right]\right\}
\label{firstomega}\\
 &+&\half\int_{x}^{x+2\pi}dy \delta f_i\delta\dot{f}_i \>.\nonumber
\end{eqnarray}

Note now that if $a +b$ is to be a singlevalued function on the circle,
then $a$ and $b$ can be expanded as
\begin{eqnarray}
a(x^+) &=& a_0 + s(t+x) +\sum_{n\ne 0} a_n e^{-
inx^+}\>,\label{24.c}\\
b(x^-) &=& b_0 + s(t-x) +\sum_{n\ne 0} b_n e^{-
inx^-}\>.\label{25.c}\end{eqnarray}
Replacing this expansion into (\ref{firstomega}) we find after some
integrations
by parts

\begin{eqnarray}
\omega = &-&\delta \frac{M}{\lambda}\delta r \nonumber\\
         &+&\int_x^{x + 2\pi} d y\left\{(\delta\pm b\delta \ln w_-
+\delta\pp a \delta \ln w_+) + \half(\delta f_i\delta \dot f_i)\right\}\>,
\label{omega}\end{eqnarray}
where $M$ is the zero mode of $a + b$, $M =\lambda(a_0 +
b_0)$.

        The symplectic form (\ref{omega}) can be brought to an even simpler
 form by making the change of variables given by

\begin{eqnarray} \pp a &=& \tilde a w_+\>, \\
\pm b &=& \tilde b w_-\>.\end{eqnarray}
In this variables the symplectic form is written as

\begin{eqnarray}
\omega = &-&\delta \frac{M}{\lambda}\,\delta r\nonumber\\
 &+&\int_x^{x+2\pi} d y \left\{(\delta \tilde b \delta w_-
+ \delta \tilde a\delta w_+) + \half(\delta f_i\delta \dot f_i)\right\}\>.
\label{lastomega}\end{eqnarray}

The constraints as well
take in these variables an specially simple form since they
are given by

\begin{eqnarray}
T_{++} &=& (\pp \tilde a - \frac{r}{2\pi}\tilde a)w_+  +
\half\sum_i\pp f_i\pp f_i\>,\label{lastTplusplus}\\
T_{--}&=&(\pm\tilde b - \frac{r}{2\pi}\tilde b)w_- +
\half\sum_i\pm f_i\pm f_i\>.\label{lastTminusminus}
\end{eqnarray}

Functions $w^+$ and $w^-$ are pure gauge, and we could set them to be
constants, so that
\be
\omega = -\delta \frac{M}{\lambda}\,\delta r
 +\half\int_x^{x+2\pi} d y \delta f_i\delta \dot f_i \>.
\label{lastom}\ee
The analysis of the reduced phase space is more complicated
then in the pure gravity case, but the fact that the symplectic form
decomposes as in eq. (\ref{lastom}) can simplify the analysis.
As in the pure case, the reduced phase
space should be determined from the analysis of the solutions.
Given the difficulties of that approach, we can say that r.p.s.
is a direct product of a subspace of $T^*(\R)$ with the phase
space for the free scalar fields.

\section{The noncompact case}

In this section the spacetime manifold will be taken as
${\rm R}\times{\rm R}$. The spacelike sections will be now noncompact and
with boundary. This fact introduces important differences with
respect to the previous case, when the manifold did not have a
boundary.

The general solution in the pure dilaton gravity case takes the form
\begin{eqnarray}
\ephi &=& -\lambda^2pm +\frac{M}{\lambda}\>,\nonumber\\
\erho &=&\frac{\pp p\pm m}{-\lambda^2pm +
\frac{M}{\lambda}}\>.\label{ncs}
\end{eqnarray}
The symplectic form is now
\begin{eqnarray}
\omega =\int_{-\infty}^{+\infty}dx{d\over dx}\left\{
\delta\frac{M}{\lambda}\delta\ln\pm m
+\delta(-\lambda^2pm)\delta\ln (p\pm m )
\right\}\>.\label{nf}
\end{eqnarray}
$\o$
must be such that it is independent of $t$ and of the hypersurface of
integration, i.e, it must be given only in terms of constants of
motion and their conjugate momenta. If one makes the standard gauge
choice $p=x^+$, $m = x^-$, the symplectic form (\ref{nf}) vanishes.
This is contrary to our expectation that the system has only one
degree of freedom, i.e. $M$. The reason why one gets a zero result
is that there is no other parameter to play the role of a conjugate
momentum to $M$. In the compact case that parameter was the monodromy
parameter $q$, which was introduced from the observation that $p$ and
$m$ do not have to be singlevalued functions on the circle.
In the non-compact case we could try to introduce an analogous parameter $q$
by means of the choice
\be p= qx^+ \quad,\quad m = {1\over q} x^- \label{ngc}\ee
However, this election cannot give a meaningful result for $\omega$
since the scaling
\be p(x^+)\to qp(x^+) \quad,\quad m (x^-) \to {1\over q} m (x^-)
\label{scal}\ee
preserves the form of the solution (\ref{ncs}).
We would get a meaningful result for $\o$
\be \o =  \d {M\over\l}\d \ln q \label{ncf}\ee
provided we evaluate the form only at one end of the real line. However, even
if we
somehow justify the evaluation of the symplectic form at one end,
this prescription does not work in the case when the matter is
present, since then eq. (\ref{omega}) becomes
\be \o = \half\int \d f_i \d \dot{f}_i \quad.\ee

Introduction of a wall, a device employed in the
thermodynamics of 2d black holes \cite{pergib}, may provide some ideas
how to proceed.
Also one could try to exploit the ambiguity of the
symplectic current under addition of a divergenceless one-form, which
in the non-compact case changes the symplectic form by a surface term.
Whichever way one chooses to proceed, the final expression should be
gauge independent and independent of the spatial surface.

On the other hand, knowledge of the symplectic form is not necessary
in order to perform the quantization on the space of the solutions. It is
only a device which simplifies the analysis of the structure of the
phase space.

After completing our work, we received a preprint
\cite{hirano} where the standard canonical
formalism is applied to the space of the solutions in the compact case
with matter. Instead of working with the symplectic form, authors of
\cite{hirano} work with the Poisson brackets. They find a canonical
transformation which maps the constraints into a quadratic form
similar to our eqs. (\ref{lastTplusplus}),(\ref{lastTminusminus}).
However, they do
not address the problem of the reduced phase space.

\v 0.5cm
\noindent{\bf Acknowledgements.}\hfil\break
M.N. is grateful to the MEC for a FPU grant. M.N. acknowledges the Imperial
College, where this paper have been written, for its hospitality. M.N. would
like to thank J. Navarro-Salas and C. Talavera for valuable comments.

\end{document}